\documentclass[a4paper,11pt]{article}
\usepackage{graphicx,rotating}
\usepackage{hyperref} 
\pdfoutput=1
\hypersetup{colorlinks,bookmarksopen,bookmarksnumbered,citecolor=verdes,
linkcolor=blus,pdfstartview=FitH,urlcolor=rossos}
\def\hhref#1{\href{http://arxiv.org/abs/#1}{#1}} 

\newcommand{\Mpl}{M_{\rm Pl}}

\usepackage{amsfonts}
\usepackage{amsmath}
\usepackage{amssymb}
\usepackage{graphicx}%
\def\npb#1#2#3{    {\it Nucl. Phys. }{B #1} (#2) #3}
\def\plb#1#2#3{    {\it Phys. Lett. }{B #1} (#2) #3}

\def\prd#1#2#3{    {\it Phys. Rev. }{D #1} (#2) #3}

\def\prep#1#2#3{   {\it Phys. Rep. }{#1} (#2) #3}
\def\prl#1#2#3{    {\it Phys. Rev. Lett. }{#1} (#2) #3}

\def\ppnp#1#2#3{   {\it Prog. Part. Nucl. Phys. }{#1} (#2) #3}

\def\zpc#1#2#3{    {\it Z. Phys. }{C #1} (#2) #3}

\def\ibid#1#2#3{   {\it ibid. }{#1} (#2) #3}

\usepackage{slashed}

\newcommand{\beq}{\begin{equation}}
\newcommand{\eeq}{\end{equation}}
\newcommand{\fig}[1]{~\ref{fig:#1}}

\newcount\Mac  \Mac=1  
\newcommand{\ifMac}[2]{\ifnum\Mac=1 #1 \else #2 \fi}
\def\putps(#1,#2)(#3,#4)#5#6{\ifnum\Mac=1 \put(#1,#2){\special{picture #5}}
\else  \put(#3,#4){\includegraphics{#6}} \fi}

\newcommand{\One}{\hbox{1\kern-.24em I}}

\newcommand{\GeV}{\,{\rm GeV}}

\newcommand{\eq}[1]{~{\rm (\ref{eq:#1})}}

\newcommand{\lascia}[1]{}
\makeatletter
\def\art{\@ifnextchar[{\eart}{\oart}}
\def\eart[#1]#2#3#4#5#6{{\rm #2}, {#3 #4} {\rm (#6) #5} [{\hhref{#1}}]}
\def\hepart[#1]#2{{\rm #2, \hhref{#1}}}
\newcommand{\oart}[5]{{\rm #1}, {#2 #3} {\rm (#5) #4}}

\newcounter{alphaequation}[equation]
\def\thealphaequation{\theequation\hbox to
0.6em{\hfil\alph{alphaequation}\hfil}}
\def\eqnsystem#1{
\def\@eqnnum{{\rm (\thealphaequation)}}
\def\@@eqncr{\let\@tempa\relax \ifcase\@eqcnt \def\@tempa{& & &} \or
  \def\@tempa{& &}\or \def\@tempa{&}\fi\@tempa
  \if@eqnsw\@eqnnum\refstepcounter{alphaequation}\fi
\global\@eqnswtrue\global\@eqcnt=0\cr}
\refstepcounter{equation} \let\@currentlabel\theequation \def\@tempb{#1}
\ifx\@tempb\empty\else\label{#1}\fi
\refstepcounter{alphaequation}
\let\@currentlabel\thealphaequation
\global\@eqnswtrue\global\@eqcnt=0 \tabskip\@centering\let\\=\@eqncr
$$\halign to \displaywidth\bgroup \@eqnsel\hskip\@centering
$\displaystyle\tabskip\z@{##}$&\global\@eqcnt\@ne
\hskip2\arraycolsep\hfil${##}$\hfil& \global\@eqcnt\tw@\hskip2\arraycolsep
$\displaystyle\tabskip\z@{##}$\hfil
\tabskip\@centering&\llap{##}\tabskip\z@\cr}

\def\endeqnsystem{\@@eqncr\egroup$$\global\@ignoretrue} \makeatother

\def\Lag{{\cal L}}

\def\circa#1{\,\raise.3ex\hbox{$#1$\kern-.75em\lower1ex\hbox{$\sim$}}\,}

\newcommand{\Mp}{M_{\rm Pl}}

\usepackage{multicol}
\usepackage{color}
\definecolor{rosso}{cmyk}{0,1,1,0.4}
\definecolor{rossos}{cmyk}{0,1,1,0.55}
\definecolor{rossoc}{cmyk}{0,1,1,0.2}
\definecolor{blu}{cmyk}{1,1,0,0.3}
\definecolor{blus}{cmyk}{1,1,0,0.6}
\definecolor{bluc}{cmyk}{1,1,0,0.1}
\definecolor{verde}{cmyk}{0.92,0,0.59,0.25}
\definecolor{verdec}{cmyk}{0.92,0,0.59,0.15}
\definecolor{verdes}{cmyk}{0.92,0,0.59,0.4}
\definecolor{grigio}{cmyk}{0,0,0,0.07}
\definecolor{rosa}{cmyk}{0,0.1,0.1,0.02}
\definecolor{rosino}{cmyk}{0,0.05,0.05,0.02}
\definecolor{rosas}{cmyk}{0,0.3,0.25,0.05}
\definecolor{celeste}{cmyk}{0.1,0,0,0.02}
\definecolor{giallino}{cmyk}{0,0,0.4,0.02}
\definecolor{rosso}{cmyk}{0,1,1,0.4}
\definecolor{rossos}{cmyk}{0,1,1,0.55}
\definecolor{rossoc}{cmyk}{0,1,1,0.2}
\definecolor{blu}{cmyk}{1,1,0,0.3}
\definecolor{bluc}{cmyk}{1,1,0,0.1}
\definecolor{blucc}{cmyk}{0.7,0.5,0,0}
\definecolor{viola}{cmyk}{0,1,0,0.6}
\definecolor{viola2}{cmyk}{0,1,0.2,0.6}
\definecolor{verde}{cmyk}{0.92,0,0.59,0.25}
\definecolor{verdec}{cmyk}{0.92,0,0.59,0.15}
\definecolor{verdes}{cmyk}{0.92,0,0.59,0.4}
\definecolor{verdino}{cmyk}{0.12,0,0.09,0.05}
\definecolor{giallo}{cmyk}{0,0,1,0}
\definecolor{gialloverde}{cmyk}{0.44,0,0.74,0}

\font\tenrsfs=rsfs10 at 11pt
\font\sevenrsfs=rsfs7
\font\fiversfs=rsfs5
\newfam\rsfsfam
\textfont\rsfsfam=\tenrsfs
\scriptfont\rsfsfam=\sevenrsfs
\scriptscriptfont\rsfsfam=\fiversfs
\def\mathscr#1{{\fam\rsfsfam\relax#1}}
\def\Lag{\mathscr{L}}

\begin{document}
\color{black}
\vspace{1.0cm}

\begin{center}
{\Huge\bf\color{rossos}Gravitational corrections to \\ [3mm]
Standard Model vacuum decay}\\
\medskip
\bigskip\color{black}\vspace{0.6cm}
{
{\large\bf Gino Isidori}$^{a,b}$,
{\large\bf Vyacheslav S. Rychkov}$^a$,
{\large\bf Alessandro Strumia}$^c$,
{\large\bf Nikolaos Tetradis}$^d$
}
\\[7mm]
{\it $^a$ Scuola Normale Superiore and INFN, Piazza dei Cavalieri 7, I-56126 Pisa, Italy}\\[3mm]
{\it $^b$ INFN, Laboratori Nazionali di Frascati, Via E.Fermi 40, I-00044 Frascati, Italy}\\[3mm]
{\it  $^c$ Dipartimento di Fisica dell'Universit{\`a} di Pisa and INFN, Italia}\\[3mm]
{\it  $^d$ Department of Physics, University of Athens, Zographou GR-15784, Athens, Greece}

\bigskip\bigskip\bigskip

{\large
\centerline{\large\bf Abstract}

\begin{quote}
We refine and update the metastability constraint on the Standard Model top and
Higgs masses, by analytically including
gravitational corrections to the vacuum decay rate.
Present best-fit ranges of the top and Higgs masses mostly lie in the narrow metastable region.
Furthermore, we show that the SM potential can be fine-tuned in order to be made suitable for inflation.
However, SM inflation results in a power spectrum of cosmological perturbations
not consistent with observations. 
\end{quote}}

\end{center}

\section{Introduction}
Assuming that the Standard Model (SM) holds up to some high energy scale
close to $M_{\rm Pl}=1.22~10^{19}\*\GeV$, present data suggest a light
Higgs mass, $m_h \sim (115-150)\GeV$.
If the Higgs is so light, radiative corrections induced
by the top Yukawa coupling can destabilize the Higgs potential and the
electroweak vacuum
becomes a false vacuum, which sooner or later decays~\cite{coleman_sc,stab,sher,IRS,stab_new,ant}.
Demanding that the SM vacuum be sufficiently long lived
with respect to the age of the universe implies a bound
on the Higgs and top masses~\cite{stab,sher,IRS,stab_new}.

In section~\ref{vacSM} we recall the peculiarities of vacuum decay within the SM
relevant for our later inclusion of gravity, which was neglected in previous analyses.
In section~\ref{vacG} we show how gravitational corrections to the vacuum decay rate~\cite{CL}
can be computed by making a perturbative expansion in the Newton constant,
and we obtain the analytic result for SM vacuum decay.


In section~\ref{infl} we show that for fine-tuned values of the
Higgs and top masses (that lie within the experimentally allowed range), 
the SM potential can be suitable for inflation.
However, the corresponding power spectrum of anisotropies  
is larger than the observed one.

\section{Vacuum decay within the Standard Model}\label{vacSM}
We recall vacuum decay within the Standard Model without gravity, and its peculiarities
relevant for our later inclusion of gravity.
The SM contains one complex scalar doublet $H$,
\beq
H = \left[ \begin{array}{c}  (h+i\eta)/\sqrt{2} \\ \chi^- \end{array} \right]~,
\eeq
with tree-level potential
\beq
\label{eq:SMV}
V=m^2 |H|^2 + \lambda |H|^4
=\frac{1}{2} m^2 h^2 + \frac{1}{4}\lambda h^4+\ldots
\eeq
where the dots stand for terms that vanish when the Goldstones $\eta,\chi^-$ are set to
zero.
With this normalization, $v=
(G_F\sqrt{2})^{-1/2}= 246.2 \GeV$, and the mass of the single physical
degree of freedom $h$ is $m^2_h=V''(h)|_{h=v}=2 \lambda v^2$. As is
well known, for $h \gg v$ the quantum corrections to $V(h)$ can be
reabsorbed in the running coupling $\lambda(\bar\mu)$, renormalized at a
scale $\bar\mu \sim h$. To a good accuracy, $V(h \gg v )\approx \lambda(h) h^4/4$ and the instability occurs 
if $\lambda$ becomes negative for some value of $h$.
For the values of $m_h$ compatible with data this occurs at scales larger than $10^5$~GeV,
suggesting that we can compute vacuum decay neglecting the quadratic term $m^2h^2/2$.

\medskip

The bounce~\cite{coleman_sc} is a solution $h(r)$ of
the Euclidean equations of motion that depends only on the radial coordinate
$r^2\equiv x_\mu x_\mu$:
\beq\label{eq:bounce}
-\partial_\mu \partial_\mu h+V'(h )
=-\frac{d^2 h }{dr^2}-\frac{3}{r}\frac{dh }{dr}+V'(h)=0~,
\eeq
and satisfies the boundary conditions
\beq
h'(0)=0~,\qquad h(\infty) = v\to 0~.
\eeq
We can perform a tree-level computation
of the tunnelling rate with a negative $\lambda<0$ renormalized at some arbitrary scale $\mu$.
In this approximation, the tree-level
bounce $h_0(r)$ can be found analytically and depends on an arbitrary scale $R$:
\begin{equation}
\label{eq:fubini}
h_0(r) = \sqrt{\frac{2}{|\lambda| }}\frac{2R}{r^2+R^2}~,\qquad S_0[h_0]
=\frac{8\pi^2}{3|\lambda|}~.
\end{equation}
At first sight, computing the decay rate among two vacua in the
approximation   $V(h)=\lambda h^4/4$ where no vacuum exists may
appear rather odd. However~\cite{IRS}, the presence of a potential barrier
around the false vacuum $h\sim 0$ is not necessary, since in quantum
field theory the bounce is not a constant field configuration, and
the energy in its gradient effectively acts as a potential barrier.
Furthermore, the decay rate does not depend on the unknown physics
that eventually stabilizes the true vacuum at $h \sim \Mp$, if the
bounce has size $R\gg 1/\Mp$: once a tunneling bubble appears,
the instability due to $V'(h(0))\neq 0$ brings $h$ down to the true
minimum with unit probability. Formally, by performing the analytic
continuation from Euclidean $r^2 = x^2 + t^2$ to Minkowskian $r^2 =
x^2 - t^2$ space-time, the evolution inside the bubble is described
by eq.\eq{fubini} with $r^2<0$: $h_0$ reaches a singularity at
$r^2=-R^2$. Indeed our potential is unbounded from
below.\footnote{~In the usual case, with a potential with two minima,
the bounce can be computed only numerically. The analytic
continuation can be done by switching $r\to ir$ in eq.\eq{bounce} at
$r<0$, and solving numerically. The qualitative behavior of the
solution can be understood by noticing that this operation is
equivalent to flipping the sign of $V$: $h$ oscillates around the
true minimum, reaching it at $r\to - \infty$, i.e. for
asymptotically large times inside the expanding bubble.} In general,
what happens inside the bubble does not affect the tunneling rate
nor the growth of the bubble: being an O(4)-invariant configuration
(i.e.\ the bounce depends only on $r$), its walls expand at the
speed of light, so that what happens in the interior cannot causally
affect the exterior.

\medskip

The arbitrary parameter $R$ appears in the expression of the 
SM bounce $h_0(r)$ because in our approximations the tree level SM
potential is scale-invariant: at this level, there is an infinite set of bounces of
varying size $R$, all with the same action $S_0[h_0]$.

Quantum corrections are the dominant effect that breaks scale invariance, and have been computed in~\cite{IRS}.
At one-loop order, the tunnelling probability in the universe space-time volume $V_U$ is then given by
\beq
\label{eq:p1L}
p = \max_R \left[p(R)\right],\qquad p(R)=\frac{V_U }{R^4}  e^{-S},
\eeq
where $S =S_0 + \Delta S_{\rm 1-loop}$ is the one-loop action:
since the bounce is not a static configuration,
corrections to both the potential part, as well as to the kinetic part of the action,
must be taken into account~\cite{IRS}.
To find the bounce configuration that extremizes $S$, it is enough to evaluate it along
the family of tree-level bounces, $h_0$ in eq.\eq{fubini}, and minimize with respect to $R$.
The result is roughly $S\approx 8\pi^2/3|\lambda(\bar\mu=1/R)|$,
i.e.\ one-loop corrections remove the tree-level ambiguity on the RGE scale $\bar\mu$ by
fixing it to be the scale $1/R$ of the bounce.
Since within the SM $\lambda(\bar\mu)$ happens to run reaching a minimal 
value at $\bar\mu \sim 10^{16-17}\GeV$, tunneling is dominated by bounces with this size.
A posteriori, this justifies having neglected the SM mass term, that gives a correction
$\Delta S \sim (mR)^2 \ll 1$ to the bounce action,
and suggests that gravity should be taken into account perturbatively.


\section{Vacuum decay with gravity}\label{vacG}
We now extend the previous computation taking into account
gravity~\cite{CL}. In our case this is a potentially
relevant effect, since gravity breaks scale-invariance and $1/R$ is just
somewhat smaller than the Planck scale. One might worry that gravity
can have dramatic effects, and that the decay rate starts to depend
on the unknown depth $V_{\rm min}$ of the true minimum of the SM
potential.\footnote{~This is what one would na\"{\i}vely guess from
the results of~\cite{CL} for the bounce action: \beq \label{eq:Sthin} S_{\rm
with~gravity} \approx S_{\rm without~gravity}/[1 + R^2 V_{\rm min}/M_{\rm
Pl}^2]^2,\eeq i.e.\ the bubble does not exist if the true minimum
has a large negative cosmological constant, e.g. $V_{\rm min}\sim -
M_{\rm Pl}^4$.
However, eq.\eq{Sthin} holds within the thin-wall approximation~\cite{CL},
not applicable when $V_{\rm min}$ is large and negative, and not applicable to the SM case we are interested in.}
This is not the case. Since the exterior geometry is the flat Minkowski space,
the generic argument given in the non-gravitational case still holds:
the bubble is an O(4)-invariant solution and its walls expands at the speed of light,
irrespectively of what happens inside.\footnote{~It is only an observer inside the bubble that experiences
a large negative cosmological constant and consequently a contraction down to a big-crunch singularity~\cite{CL},
instead than  an expanding bubble.}

\medskip

We recall from ~\cite{CL} the basic formalism needed for a quantitative analysis.
We assume an Euclidean spherically symmetric geometry,
$ds^2 = dr^2 + \rho(r)^2 d\Omega^2$,
where $d\Omega$ is the volume element of the unit 3-sphere.
The Einstein-Higgs action
\beq S =  \int d^4x \sqrt{g} \left[ \frac{(\partial_\mu h)(\partial^\mu h)}{2} + V(h)  -
\frac{\mathscr{R}}{2\kappa}\right],\label{eq:EH}
\eeq
where $\kappa = 8\pi G$  and $G = 1/\Mp^2$ with $M_{\rm Pl}=1.22~10^{19}\,{\rm GeV}$, simplifies to
\beq S = 2\pi^2 \int dr\left[\rho^3( \frac{h^{\prime 2}}{2} + V )+ \frac{3}{\kappa}
(\rho^2 \rho'' + \rho\rho^{\prime 2} - \rho)\right],
\eeq
where $'$ denotes $d/dr$.  The equations of motion are
\beq h'' + 3\frac{\rho'}{\rho} h' = \frac{dV}{dh},\qquad
\rho^{\prime 2} = 1 + \frac{\kappa}{3}\rho^2 (\frac{h^{\prime 2}}{2} - V ).\eeq

\begin{figure}[t]
\begin{center}
\includegraphics[width=0.6\textwidth]{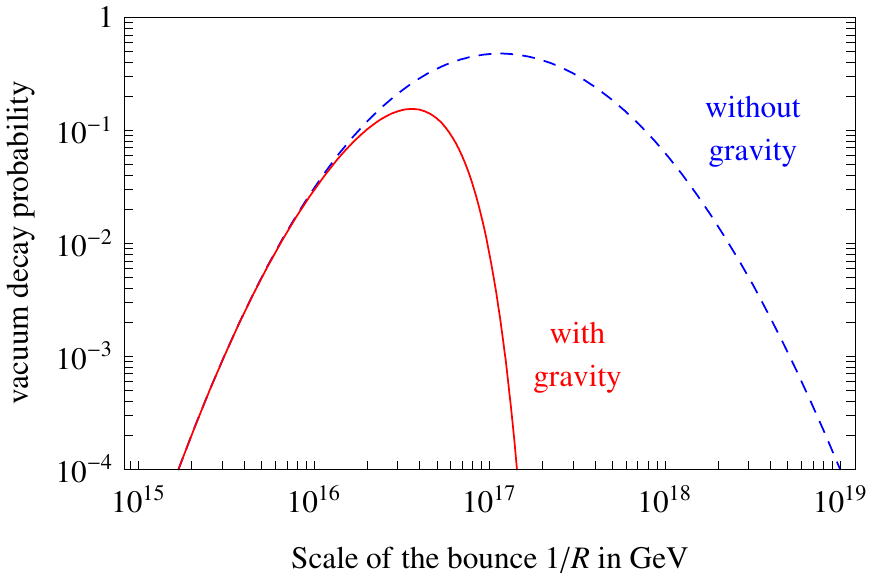}
\caption{\em\label{fig:probR} Probability $p(R)$ that the SM vacuum decayed so far for
$m_h=115\GeV$, $m_t = 174.4\GeV$, $\alpha_3(M_Z)=0.118$,
due to bounces with size $R$,
without including gravitational effects (dashed curve~\cite{IRS}) and including gravitational effects (continuous line).
The correction is relevant only at $1/R\circa{>}10^{17}\GeV$.
Uncertainties due to higher-order corrections are not shown.}
\end{center}
\end{figure}

\begin{figure}[t]
\begin{center}
\includegraphics[width=0.8\textwidth]{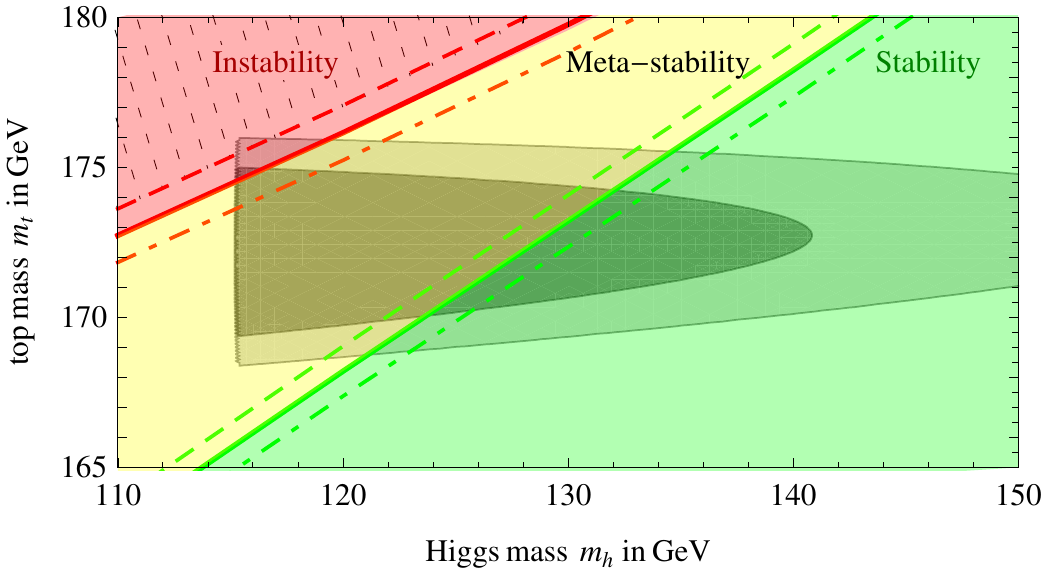}
\caption{\em\label{fig:deadoralive} Metastability region of the
Standard Model vacuum in the $(m_h,m_t)$ plane, for
$\alpha_s(m_Z)=0.118$ (solid curves). Dashed and dot-dashed curves
are obtained for $\alpha_s(m_Z)=0.118\pm 0.002$.
The shaded half-ellipses indicates the experimental range for $m_t$ and $m_h$ at $68\%$ and $90\%$ confidence level.
Sub-leading effects could shift the bounds by $\pm 2\GeV$ in $m_t$.}
\end{center}
\end{figure}

We can analytically include the effect of gravity, assuming $ R M_{\rm Pl}\gg 1$, by performing a 
leading-order 
expansion in  the gravitational
coupling $\kappa$:
\beq h(r) = h_0(r) + \kappa h_1(r)+{\cal O}(\kappa^2)  ,\qquad \rho(r) = r + \kappa \rho_1(r)+{\cal O}(\kappa^2) .\eeq
The action is
\beq \label{eq:Gexp}
S = S_0 + 6\pi^2\kappa \int dr~  \left[r^2 \rho_1 \left( \frac{h_0^{\prime 2}}{2} + V(h_0) \right)
+(r \rho_1^{\prime 2} + 2 \rho_1 \rho_1' +  2 \rho_1r \rho_1'')\right]+
{\cal O}(\kappa^2).\eeq
We have taken into account that many terms in the expansion vanish either because
the integrand is a total derivative (e.g.\  the negative power $1/\kappa$ in eq.\eq{EH} is just apparent)
or thanks to the equations of motion.
Indeed $h_1$ does not appear in eq.\eq{Gexp} because we are  functionally expanding around the extremum $h_0$ of the non-gravitational action,  so that the first functional derivatives vanish thanks to the equations of motion.
So, we only need to compute $\rho_1$: its equation of motion is \beq
\rho'_1 = \frac{1}{6}r^2 \left(\frac{h_0^{\prime 2}}{2} - V(h_0)
\right).\eeq Inserting it into eq.\eq{Gexp} completes the
computation of gravitational corrections to leading-order in $\kappa$. We
notice that the first term in eq.\eq{Gexp}, which is linear in
$\rho_1$, contributes $-2$ times the last purely gravitational term
in eq.\eq{Gexp}, which is quadratic in $\rho_1$. This happens
because $S$ must have an extremum at $c=1$ under the variation
$\rho_1(r) \to c \rho_1(r)$. The discussion is so far general, and
by choosing toy potentials we verified that eq.\eq{Gexp} agrees with
the full numerical result.

\section{Vacuum decay with gravity in the Standard Model}\label{infl}
Going to the SM case, using the analytic expression of eq.\eq{fubini} for the bounce $h_0$,
we can perform all integrations analytically finding
\beq 
\label{eq:SSM}
 S = \frac{8\pi^2}{3|\lambda|}+\Delta S_{\rm 1-loop} +\Delta S_{\rm gravity}, 
\qquad \Delta S_{\rm gravity}=\frac{256\pi^3}{45 (R M_{\rm Pl} \lambda)^2}
\eeq
where $\Delta S_{\rm gravity}$ 
is the gravitational correction and 
$\Delta S_{\rm 1-loop}$ the one-loop correction, given in eq.~(3.3) of~\cite{IRS}.
Eq.\eq{p1L} gives the tunneling probability $p(R)$.


Fig.\fig{probR} shows an example of the relevance of gravitational
corrections. We checked that the leading-order approximation agrees
with the result of a full numerical computation: eq.\eq{SSM}
correctly approximates the action of the true bounce, and  the true
bounce $h(r)$ is correctly approximated by $h_0(r)$ with the value
of $R$ that minimizes $S$.\footnote{~Here we comment about the comparison between the analytic 
result in eq.~(\ref{eq:SSM}) and the full numerical computation.
With a typical potential this is a straightforward procedure:
the bounce is determined numerically as a compromise between 
classical solutions which under-shoot and over-shoot the true bounce at large $r$.
With a potential close to $h^4$, finding the bounce numerically is more
involved: with this potential classical solutions necessarily go
to zero at large $r$; however, they generically oscillate to zero as $1/r$ 
giving a divergent action. The special feature of $h_0(r)$ 
is that it vanishes as $1/r^2$ giving a finite action.
The true bounce should maintain this behavior.
In practice, this is achieved imposing a vanishing 
difference between $h_0(r)$ and the numerical bounce. 
The advantage of our analytic approximation based on the set of candidate bounces
$h_0(r)$ with different values of $R$ is that ill-behaved never enter the computation.
}

Fig.\fig{deadoralive} shows the regions in the $(m_h,m_t)$ plane where the SM vacuum is stable, meta-stable or too unstable.
Gravitational corrections only induce a {\em minor shift} on the `instability' border,
less relevant than  present experimental and theoretical (higher-order) uncertainties.
The ellipses truncated at $m_h=115\GeV$ are the best-fit values for the top and Higgs masses,
from our up-to-date global fit of precision data, that includes
the latest direct measurement of  the top mass, $m_t = (170.9\pm1.8)\GeV$~\cite{mt}.
Present data and computations indicate that we do not live in the unstable region
(such that the SM can be valid up to the Planck scale),
but increased accuracy is needed to determine if we live in the stable or in the small meta-stable region.

\medskip

Adding to the SM action possible dimension-6 non-renormalizable
operators suppressed by the Planck scale would give similar
corrections to the bounce action. In particular, adding to the
SM Lagrangian the operators
\beq
 \Delta\Lag_6=\frac{1}{M_{\mathrm{Pl}}^{2}} \left(
- \xi M_{\mathrm{Pl}}^{2} \mathscr{R}|H|^{2}+c_{1}  \frac{|H|^{6} }{3!} +c_{2}|H|^{2}|D_\mu H|^{2} \right),
\eeq
where $\xi$ and $c_{1,2}$ are unknown dimensionless coefficients,
gives the following correction 
\begin{equation}
\Delta S^\prime_{\rm gravity} = \frac{8\pi^{2}}{15(M_{\mathrm{Pl}}R\lambda)^{2}} \left( 128\pi\xi
+ \frac{c_{1}}{|\lambda|}+4c_{2} \right),\label{eq:other}%
\end{equation}
which can be 
comparable to the model-independent gravitational effect computed in eq.~\eq{SSM}.

The values of the coefficients $\xi$ and $c_{1,2}$ change 
under field redefinitions and only their linear combination 
entering (\ref{eq:other}) is physical. Indeed, under $H\rightarrow H(1+a|H|^{2}%
/M_{\mathrm{Pl}}^{2})$ we have\footnote{~We do not distinguish
between $|H^{\dagger}D_\mu H|^{2}$ and $|H|^{2}|D_\mu H|^{2}$
since these operators coincide on the configurations
$H=(h/\sqrt{2},0)$ we are interested in.} $\delta c_{1}=24\lambda
a,$ $\delta c_{2}=6a$ and $\delta\xi=0$; this transformation can be used
to set $c_{2}\rightarrow0$. Under the Weyl
transformation of the metric $g_{\mu\nu}\rightarrow g_{\mu\nu}(1+a|H|^{2}%
/M_{\mathrm{Pl}}^{2})$ we have $\delta\xi=a/16\pi$, $ \delta
c_{1}=12a\lambda$, $\delta c_{2}=a$; this transformation can be used
to set $\xi\rightarrow0$. Both these field redefinitions
leave $\Delta S^\prime_{\rm gravity}$ invariant. 

To estimate the magnitude of $\Delta S^\prime_{\rm gravity}$ we can thus restrict
the attention to only one of the three operators in $\Delta \Lag_6$
(we choose the $|H|^{6}$ term), and estimate its coupling using na\"{\i}ve 
dimensional analysis. At one loop, graviton exchanges
generate the $|H|^{6}$ operator with $c_{1}\sim
g_{\mathrm{s}}^{4}/\pi$ as well as the $\lambda|H|^{4}$ operator
with coefficient $\lambda\sim g_{\mathrm{s}}^{4}/\pi^{2}$. Here
$g_{\mathrm{s}}$ is an unknown coefficient which determines if
quantum gravity is weakly or strongly coupled, with strong coupling
corresponding to $g_{\mathrm{s}}\sim\pi^{2}$. One might therefore
argue that $c_{1}\sim \lambda\pi$, which implies 
$\Delta S_{\rm gravity}^\prime \sim \Delta S_{\rm gravity}$.

\begin{figure}[t]
$$\includegraphics[width=13cm]{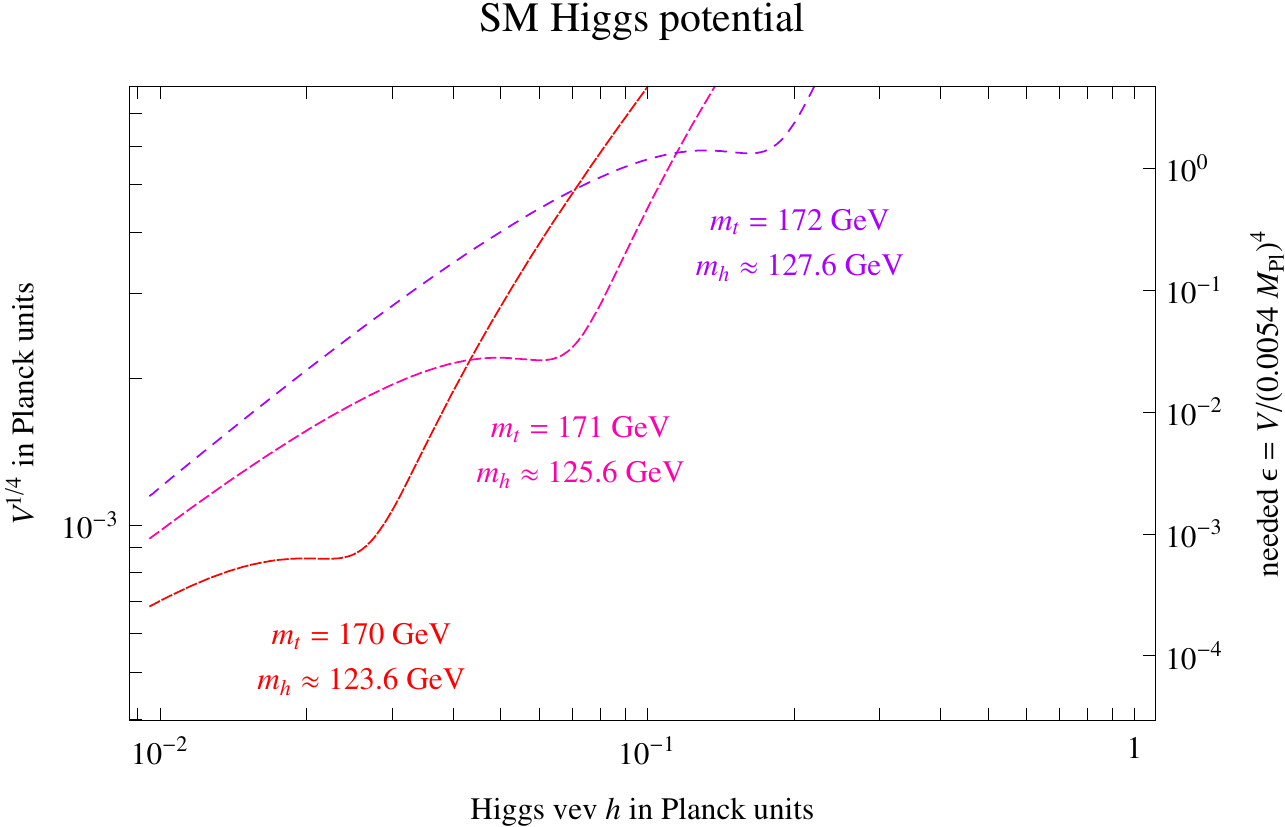}$$
\caption[X]{\label{fig:VSM}\em Examples of fine-tuned SM potentials that might allow inflation.
The right handed axis shows the value of the slow-roll parameter $\varepsilon$ that would give the observed amount of anisotropies.
}
\end{figure}

\section{Inflation within the Standard Model?}
For $m_t \approx 173 \GeV$ and $m_h \approx 130\GeV$
(i.e.\ within the experimentally allowed region)
both the quartic Higgs coupling $\lambda$ and its $\beta$-function happen to vanish,
at some RGE scale around $M_{\rm Pl}$.
Is this just a coincidence, or this boundary condition carries some message?
Some speculations about this fact have been presented in~\cite{Nielsen}. Here we explore 
a different aspect, namely a possible connection with inflation.

%

The quasi-vanishing of both $\lambda$ and $\beta(\lambda)$ allows to have a
quasi-flat Higgs potential at $h\sim M_{\rm Pl}$, suitable for inflation.
Indeed, we can approximate the RGE running of $\lambda$ as
 \beq \lambda(\mu \sim h_0)\simeq \lambda_{\rm min} + \frac{\gamma}{(4\pi)^4}\ln^2\frac{\mu}{h_0}\eeq
around the special value $h_0$ where $\lambda$ reaches its minimal value $\lambda_{\rm min}$.
The constant $\gamma$ is related to $\beta(\beta(\lambda))$
and has the numerical value $\gamma\approx 0.6$ within the SM.
The first and second derivatives of the SM potential $V \simeq \lambda(h) h^4/4$
vanish at $h=h_*\equiv h_0e^{-1/4}$ if $\lambda_{\rm min} = \gamma/4096\pi^4$,
such that the slow-roll parameters
$$ \varepsilon\equiv
\frac{M_{\rm Pl}^2}{16\pi} \left(\frac{V'}{V}\right)^2,\qquad
\eta\equiv
\frac{M_{\rm Pl}^2}{8\pi} \frac{V''}{V}~
$$
vanish, allowing for inflation.

The lack of convincing natural models for inflation might indicate that it happens when
scalar fields,  fluctuating along some vast `landscape' potential generically unsuitable for inflation,
encounter a small portion of the potential which  accidentally is flat enough.
This is what might happen within the SM.
This potential is illustrated in fig.\fig{VSM},
where we do not show the uncertainty due to higher-order corrections,
which effectively amounts to a $\pm 2\GeV$ uncertainty in $m_t$.
Can this SM potential be responsible for inflation {\em and} the
generation of anisotropies $\delta \rho/\rho$? The answer is: not
both. The basic problem is that the requirement of having enough
$e$-folds of inflation, \beq N =2\sqrt{\pi} \int
\frac{dh/\Mpl}{\sqrt{\varepsilon}}\approx 60,\eeq can be met with a
small enough $\varepsilon$, but this conflicts with the requirement
that quantum fluctuation of the Higgs inflaton should also generate the observed power spectrum of anisotropies, $\delta \rho/\rho
\sim 10^{-5}$, i.e.\ \beq \frac{V}{\varepsilon}\approx (0.0054
\Mpl)^4.\eeq Indeed the height $V$ of the SM potential in its flat
region is predicted and cannot be arbitrarily adjusted to be as low
as needed. This result can be understood by either doing explicit
computations with the approximated potential $\lambda(h) h^4/4$, or
by looking at the sample SM potentials plotted in fig.\fig{VSM}. For
a top mass within the observed range, the plateau is at values of
$h$ and $V^{1/4}$ which are are somewhat below the Planck scale, but
$\delta\rho/\rho$ at $N\approx 60$ comes out larger than the
observed value. Successful inflation and successful generation of
anisotropies would be obtained if for some unknown reason the
potential would remain flat from $h\sim h_*$ up to $h\sim M_{\rm
Pl}$.

\begin{figure}[t]
$$\includegraphics[width=0.7\textwidth]{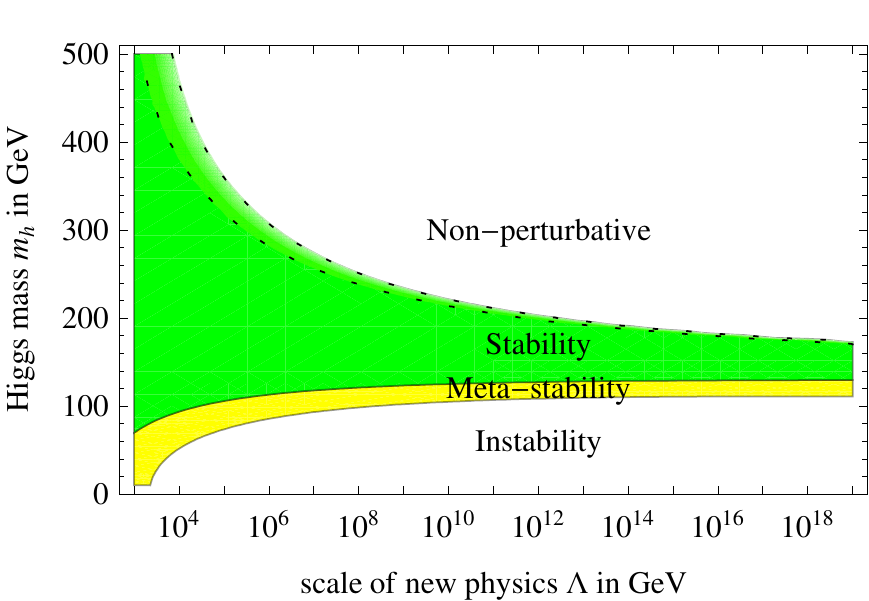}$$
\caption[X]{\label{fig:Lambda}\em 
Bounds on the Higgs mass derived by the conditions of absolute stability (lower bound), 
sufficient metastability (yellow region) and 
perturbativity (upper dotted lines, derived by the conditions $\lambda < 3,6$),
as function of the scale of validity of the SM.
This plot assumes $m_t=173\GeV$ and $\alpha_3(M_Z)=0.118$.
}
\end{figure}

\section{Conclusions}
In this paper we have refined and updated the metastability constraint
on the Higgs mass, assuming the validity of the Standard Model up to the highest possible
energy scale, $\Lambda \approx M_{\rm Pl}$.
In particular, we have taken into account gravitational corrections,
which were neglected in previous analyses. These corrections 
turn out to be small and calculable in the phenomenologically 
interesting region of $m_h$ close to its experimental lower bound.
The updated constraints in the $(m_h,m_t)$ plane are reported in fig.~\ref{fig:deadoralive}.
Among all possible values, the Higgs mass seems to lie in the
narrow region which allows the SM to be a consistent theory up to very high energy scales,
with a perturbative coupling and a stable or sufficient long-lived vacuum.
Fig.~\ref{fig:Lambda} illustrates the constraints on the Higgs mass as function of $\Lambda$,
and shows that the (meta)stability constraints do not depend on $\Lambda$ when it is around the Planck scale.


We have also shown that the SM potential can be fine-tuned in order to be made 
suitable for inflation. However, the resulting power spectrum 
of anisotropies is larger than the observed one.

\paragraph{Acknowledgements}
We thank Paolo Creminelli and Enrico Trincherini for useful discussions.

\end{document}